\documentclass[aps,12pt,preprint,groupedaddress,showpacs,reprintnumbers,amsmath,amssymb,floatfix]{revtex4}
\usepackage{epsfig}
\usepackage{longtable}
\usepackage{graphicx}
\usepackage{lscape}

\usepackage{epsf}
\usepackage{epsfig}
\usepackage{amsmath,amsfonts,amsthm, amssymb,latexsym}
\usepackage{enumerate}
\usepackage{color}
\newcommand{\IR}[1] {\textcolor{red}{#1}}

\begin{document}
\title{Proton and neutron density distributions at supranormal density in low- and medium-energy heavy-ion collisions.}
\date{\today}
\author{J.~R.~Stone}
\affiliation {Department of Physics and Astronomy, University of Tennessee, Knoxville, TN 37996, USA}
\affiliation{Department of Physics, University of Oxford, Oxford, OX1 3PU, UK}
\author{P.~Danielewicz}
\affiliation{National Superconducting Cyclotron Laboratory, Michigan State University, East Lansing, MI 48824, USA}
\author{Y.~Iwata}
\affiliation{Institute of Innovative Research, Tokyo Institute of Technology, Meguro, Tokyo 152-8550, Japan}

\begin{abstract}

{\bf{Background:}}  The distribution of protons and neutrons in the matter created in heavy-ion collisions is one of the main points of interest for the collision physics, especially at supranormal densities.   We report results of the first systematic simulation of proton and neutron density distributions in central heavy-ion collisions within the beam energy  range of $ E_{\rm beam} \leq 800 \, \text{MeV/nucl}$. The symmetric  $^\text{40}$Ca +$^\text{40}$Ca,  $^\text{48}$Ca +$^\text{48}$Ca, $^\text{100}$Sn +$^\text{100}$Sn and $^\text{120}$Sn + $^\text{120}$Sn and asymmetric  $^\text{40}$Ca +$^\text{48}$Ca and $^\text{100}$Sn +$^\text{120}$Sn systems were chosen for the simulations.

{\bf{Method:}} The Boltzmann-Uhlenbeck-Uehling (pBUU) transport model with four empirical models for the density dependence of the symmetry energy was employed.  Results of simulations using pure Vlasov dynamics were added for completeness. In addition, the Time Dependent Hartree Fock (TDHF) model, with the SV-bas Skyrme interaction, was used to model the heavy-ion collisions at $E_{\rm beam} \leq 40 \, \text{MeV/nucl}$.  Maximum proton and neutron densities $\rho_{\rm p}^{\rm max}$ and $\rho_{\rm n}^{\rm max}$, reached in the course of a collision, were determined from the time evolution of  $\rho_{\rm p}$ and  $\rho_{\rm n}$.

{\bf{Results:}} The highest total densities predicted at $E_{\rm beam} = 800 \, \text{MeV/nucl}$ were of the order of $\sim 2.5 \, \rho_{\rm 0}$ ($\rho_{\rm 0} = 0.16 \, \text{fm}^{\rm -3}$) for both Sn and Ca systems. They were found to be only weakly dependent on the initial conditions,  beam energy, system size and a model of the symmetry energy.  The proton-neutron asymmetry $\delta = (\rho_{\rm n}^{\rm max} - \rho_{\rm p}^{\rm max})/ (\rho_{\rm n}^{\rm max} + \rho_{\rm p}^{\rm max})$ at maximum density does depend, though, on these parameters. The highest value of $\delta$ found in all systems and at all investigated beam energies was  $\sim$ 0.17.

{\bf{Summary:}} We report the first systematic simulation of proton and neutron densities in matter produced in heavy-ion collisions of Ca and Sn systems at beam energies below 800 MeV/nucl using pBUU and TDHF models. We find limits on the maximum proton and neutron densities and the related proton-neutron asymmetry $\delta$ as a function of the initial state, beam energy, system size and a symmetry energy model. While the maximum densities are almost independent of these parameters, our simulation reveals, for the first time, their subtle impact on the proton-neutron asymmetry. Most importantly, we find that variations in the proton-neutron asymmetry at maximum densities are related at most at 50\% level to the details in the symmetry energy at supranormal density.  The reminder is due to the details in the symmetry energy at subnormal densities and its impact on proton and neutron distributions in the initial state. This result puts to forefront the need of a proper initialization of the nuclei in the simulation, but also brings up the question of microscopy, such as shell effects, that affect initial proton and neutron densities, but cannot be consistently incorporated into semiclassical transport models.

\end{abstract}

\maketitle

\section{Introduction}
\label{sec:intr}
The proton and neutron density distributions generated in the course of a heavy-ion collision, especially at a total density exceeding its normal value ($\rho_{\rm 0}$ = 0.16 fm$^{\rm -3}$), are among  the main points of interest of theorists, experimentalists and observers, within the communities of low and medium-energy nuclear physics and of astrophysics of compact objects. The relative asymmetry of these distributions pertains to the determination of the Equation of State (EOS) of dense matter.
The EOS is usually defined by the dependence of energy of per nucleon upon density and temperature. In the limit of zero temperature, following the semi-empirical mass formula \cite{vonW1935,bethe1936,reinhard2006}, the energy per particle can be expanded in terms of the proton-neutron asymmetry  $\delta = (\rho_{\rm n} - \rho_{\rm p})/\rho$, the local relative difference between neutron and proton densities:
\begin{equation}
\mathcal{E}(\rho,\delta)=\mathcal{E}(\rho,\delta=0)+S(\rho) \, \delta^{\rm 2} \, .
\label{eq:1}
\end{equation}
Here, $\mathcal{E}(\rho,\delta=0)$ is the energy per particle of symmetric nuclear matter, $S(\rho)$ is the symmetry energy. It  can be approximated as the difference between the energy per particle for symmetric and pure neutron matter at the same total density:
\begin{equation}
S(\rho) \equiv \frac{1}{2}\left. \frac{\partial^2\mathcal{E}(\rho,\delta)}{\partial\delta^2}\right|_{\rm \delta = 0} \approx \mathcal{E}(\rho,\delta = 1) - \mathcal{E}(\rho,\delta = 0) \, .
\end{equation}
 As the Pauli principle influences the symmetry energy, it is common to subdivide S($\rho$) into a kinetic  (no interacting)  Fermi-gas contribution $S_\text{kin}$ and an interaction contribution $S_\text{int}$, shifting the focus to $S_\text{int}$:
\begin{equation}
\label{eq:3}
S(\rho) = S_\text{kin}(\rho) + S_\text{int}(\rho) = 12.3 \, \text{MeV} \, \Big( \frac{\rho}{\rho_0}   \Big)^{2/3}  + S_\text{int}(\rho) \, .
\end{equation}
In describing a heavy-ion collision, the main uncertainties are changes in the total density $\rho$ = $\rho_{\rm n} + \rho_{\rm p}$, asymmetry $\delta$ and the magnitude and density dependence of the symmetry energy S($\rho$).

The properties of the symmetry energy provides an important clue as to why we have comparable numbers of protons and neutrons in nuclei and how the competition between the \textit{pp, pn} and \textit{nn} nucleon-nucleon interactions in nuclear environment leads to stability of nuclear systems observed in nature. No theory exists at present  that would offer a fundamental understanding of the symmetry energy and its properties. For nuclei, the value of the symmetry energy at average nuclear density is reasonably well constrained by masses of heavy nuclei and other data, such as the emission of light clusters in heavy-ion collisions~\cite{natowicz2013}. However, the value and the slope of the symmetry energy, $L = 3\rho_0 \, \partial S(\rho)/\partial (\rho)|_{\rho_0}$, and their correlation even at the normal density in symmetric nuclear matter are not known with certainty  (for recent reviews cf.~\cite{lynch2009,tsang2012,horowitz2014}). Efforts to broaden the range of densities for symmetry energy studies under laboratory conditions include pursuing observables in experiments with heavy-ion collisions and comparing them to predictions of transport models \cite{tsang2009}, such as the observables quantifying fragmentation~\cite{ono2004} and isospin diffusion in peripheral asymmetric systems~\cite{tsang2004}. Density dependence of the symmetry energy is a necessary input to these models and is provided in various empirical forms. The validity of the assumptions is judged indirectly by comparison of the outcome of transport models with experiment.

We explore the maximal densities reached separately for neutrons and protons in simulations of central heavy-ion collisions at non-relativistic and moderately relativistic energies. The asymmetry $\delta$ in high density matter and its correlation with total density are not open to direct measurement and have to remain model dependent quantities. Depending on the type of collision simulation, either single-particle wavefunctions or phase-space distributions are dynamic quantities which are followed.  Proton and neutron density profiles as a function of time of duration of an individual collision event can be simulated as a function of beam energy. The maximum proton and neutron densities provide a practical upper limit on the high-density value of $\delta$ for each case. However, it is important to realize that in a target-projectile asymmetric collision the maximal densities and maximal asymmetry may not occur at the same place and time.

 In this work we consider central collisions in the systems  $^\text{40}$Ca +$^\text{40}$Ca, $^\text{40}$Ca +$^\text{48}$Ca, $^\text{48}$Ca +$^\text{48}$Ca, $^\text{100}$Sn +$^\text{100}$Sn, $^\text{100}$Sn +$^\text{120}$Sn and $^\text{120}$Sn +$^\text{120}$Sn within the beam energy range of (0$-$800$)\, \text{MeV/nucleon}$. Calcium and tin isotopes are often chosen in experimental studies. Although the $^\text{100}$Sn +$^\text{100}$Sn system is not likely to be investigated experimentally in the near future, we include that system for completeness, as a reference. Note that the ratio of the number of neutrons to the total number of nucleons in targets and projectiles involved in the colliding systems is rather similar, being 0.5 ($^\text{40}$Ca and $^\text{100}$Sn) and 0.58 ($^\text{48}$Ca and $^\text{120}$Sn).

The paper is organized as follows. The numerical methods and their physical basis are explained in Sec.~\ref{sec:calmeth}. The results and discussion are presented in Sec.~\ref{sec:res}. Conclusions and outlook form the content of Sec.~\ref{sec:concl}.

\section{Dynamic Models}
\label{sec:calmeth}
Since we wish the simulation in this paper to provide benchmark values of proton-neutron asymmetry in heavy-ion collisions at low and medium beam energies, we give in this section, for benefit of a wider audience, a brief survey of the physics of the collisions and the models used, together with a summary of the parameters these models depend on. The physical basis of the models should be kept in mind when judging the results and may be helpful in finding ways for their future refinement.

Heavy-ion collision dynamics changes with the incident beam energy as different physical aspects play a dominant role in the energy range of $\sim 40$--$800 \, \text{MeV/nucleon}$ compared with the lower beam energies.  We discuss the Boltzmann-Uhlenbeck-Uehling (pBUU) approach, used primarily for medium-to-high beam energies, in Sec.~\ref{sec:pBUU}, and the Time-Dependent-Hartree-Fock (TDHF) method, valid only for lower beam energy collisions, in Sec.~\ref{sec:tdhf}, reflecting different physics in the low energy region .

The simulation methods employed in this work are single-particle in nature, potentially underestimating the role of fluctuations, unlike other approaches to modeling heavy-ion collisions such  as Quantum Molecular Dynamics (QMD) \cite{aichelin1991} and Antisymmetrized Dynamics (AMD) \cite{ono2004}. These methods use  Gaussian-type packets to model the particle motion and predict observables. The nature of fluctuations at high local excitation energies is becoming similar to the fluctuations in cascade models \cite{bertsch1988}. However, at lower energies, the nature and magnitude of fluctuations is not well established. In the present work we are interested in the maxima of statistically averaged densities reached during the compression stage of the collision. At this stage the fluctuations are short-lived and do not significantly affect the density evolution.  Thus single-particle approaches such as pBUU and TDHF are fully justified.

In the above context we should mention the efforts to bridge between the low-energy TDHF and high-energy Boltzmann approaches, by either supplementing the TDHF equations with an equation for orbital occupation containing a collision integral \cite{WongTang78,Ayik00} or by complementing the equation for one-body density matrix, implicit in TDHF approach, with one for two-body matrix \cite{WangCassing85}.  Realistic calculations for those approaches are limited so far to light systems and low energies~\cite{GongTohyamaRandrup90,TohyamaUmar02,TohyamaUmar02b}.

\subsection{The pBUU transport model}
\label{sec:pBUU}
Starting the discussion with the higher energy region, central collisions of heavy ions are laboratory means of bringing nuclear matter to supranormal densities over spatial regions that are large as compared to the range of the nuclear force, for times long enough to allow passage of a signal over the region. In a central collision, moving with the N-N center of mass (longitudinally), a portion of the initial kinetic energy is converted into nuclear compression energy. There will also be a lateral motion during the collision.  Particle densities higher than in ordinary nuclei can be reached at beam energies of a few hundred MeV/nucleon.

As the beam energy per nucleon increases and becomes comparable with the Fermi energy, elastic  N-N collisions begin to contribute to the compression within the system. With a~reduced role of the Pauli principle, as the Fermi spheres melt down with the occupations turning to small fractions, the N-N collisions dissipate the longitudinal motion of nucleons within the overlap region of the colliding nuclei, slowing down the passage of the nucleons across that region and causing a density pile-up when additional nucleons flow in. At even higher energies, the colliding nuclei become Lorentz contracted to some degree, before the overlap region develops. Simultaneously, the N-N collision cross sections become forward peaked, allowing the longitudinal motion to persist longer during the reaction.  In theory, the highest compression that could be reached in the N-N collision-dominated regime is given by a solution of the hydrodynamic Rankine-Hugoniot equation. This equation is applicable provided the system is large enough to allow the matter in the overlap region to reach complete equilibrium, erasing any memory of the original motion along the beam axis.  However, according to transport simulations, the Rankine-Hugoniot limit is not reached even for heaviest nuclei, independent of the beam energy \cite{danielewicz1995}.

The present numerical simulations of heavy-ion collisions in the higher-energy regime are based on solution of the semiclassical Boltzmann-Uhlenbeck-Uehling equations using the code of Danielewicz  {\it et al.}, termed recently pBUU \cite{danielewicz2000}.  The names of Uhlenbeck and Uehling are often associated with the transport equation set for heavy-ion collisions, as the two authors were the first to modify the original classical transport Boltzmann equation to include the final state Pauli suppression and Bose-enhancement factors~\cite{uehling1933}. The pBUU model is formulated within relativistic Landau theory \cite{baym1976} in which the state of a system of particles is completely specified when the phase space (Wigner) distribution functions $f_{\rm X}$ of the particles are given.  Here, $f_{\rm X}({\pmb p}, {\pmb r}, t)/(2\pi)^3 \, $ represents the phase-space probability density for finding a particle of type $X$, at time $t$  and position ${\pmb r}$, with momentum ${\pmb p}$ and a specific spin direction. Local density of particles $X$ is determined from $\rho_X ({\pmb r},t) = \frac{g_X}{(2\pi)^3} \int \text{d}{\pmb p} \, f_X({\pmb p},{\pmb r},t)$, where $g_X$ is spin degeneracy.

The distributions $f_{\rm X}({\pmb p}, {\pmb r}, t)$ satisfy a set of the Boltzmann equations \cite{danielewicz2000} (we omit here the UU for simplicity)
\begin{equation}
\frac{\partial f_X}{\partial t}+\frac{\partial\epsilon_X}{\partial {\pmb p}}\frac{\partial f_X}{\partial {\pmb r}}-\frac{\partial\epsilon_X}{\partial {\pmb r}}\frac{\partial f_X}{\partial {\pmb p}} = I_X
\label{eq:boltz}
\end{equation}
where the single-particle energies $\epsilon_X$  are derivatives of the total energy $E$ of the system
\begin{equation}
\epsilon_X({\pmb p},{\pmb r},t)=\frac{(2\pi)^3}{g_X}\frac{\partial E}{\partial f_X({\pmb p},{\pmb r},t)}.
\end{equation}
In Eq.~\eqref{eq:boltz} ${\pmb v}_X = \partial \epsilon_X/\partial {\pmb p}$ is the velocity of a particle X and ${\pmb F}_X = - \partial\epsilon_X/\partial {\pmb r}$ is the force acting on the particle. The terms on the l.h.s.\ of Eq.~\eqref{eq:boltz} account for the changes in $f_X$ due to the motion of particles in the average mean field produced by other particles. On the r.h.s.,  the collision integral $I_X$ that incorporates cross-sections and N-N collision rates~\cite{danielewicz2000} governs the modifications of the distributions $f_{\rm X}({\pmb p}, {\pmb r}, t)$ due to elastic and inelastic N-N collisions and decays caused by short-range residual interactions beyond the mean-field.  The inelastic processes, giving rise to resonances and pion production, play a significant role only at beam energies close to 800~MeV/nucleon and their effect is marginal below 400~MeV/nucleon. We employ in-medium cross sections adjusted so that the collision radius does not exceed the typical distance between nucleons in the medium \cite{danielewicz2009a}.  These cross sections adequately describe stopping observables from heavy-ion collisions, such as linear momentum transfer~\cite{danielewicz2002a}, unlike the free cross sections \cite{danielewicz1995,barker2009}. Details on the collision integrals can be found in \cite{danielewicz2000} and~\cite{danielewicz2002a} and in references therein. At low excitation energies, the N-N collisions are suppressed and the Boltzmann equations reduces to their Vlasov form~\cite{danielewicz2000}. Comparison of the pure Vlasov dynamics to that in TDHF were carried out in the past by Tang {\it et al.}~\cite{tang1981} in one dimension and by Wong~\cite{wong1982} in three dimensions.

Within the pBUU framework the collision dynamics in a mean-field approximation is determined by the dependence of the total energy $E$ on the phase-space distributions $f_X$~\cite{danielewicz2000}. The energy $E$ consists of the covariant volume term and non-covariant gradient correction~$E_{\rm 1}$, and  isospin $E_{\rm T}$ and Coulomb $E_{\rm Coul}$ terms, all defined in the local rest frame where the nucleon flux vanishes:
\begin{equation}
E = \int \text{d} {\pmb r} \, \tilde{e} + E_1 + E_T + E_\text{Coul} \,
\label{eq:funct}
\end{equation}
where $\tilde{e}$ is the volume energy density,
\begin{eqnarray}
E_1 & = & \frac{a_1}{2 \rho_0} \int \text{d} {\pmb r} \, \big(\nabla \rho \big)^2 \, ,\\[1ex]
E_T & = & 4 \int \text{d} {\pmb r} \, \frac{S_\text{int}(\rho)}{\rho} \, \rho_T^2 \, ,\\[1ex]
E_\text{Coul} & = & \frac{1}{4 \pi \epsilon_0} \int \text{d} {\pmb r} \, \text{d} {\pmb r}' \, \frac{\rho_\text{ch}({\pmb r}) \, \rho_\text{ch}({\pmb r}')}{|{\pmb r} - {\pmb r}'|} \, .
\end{eqnarray}

The expression for the energy density functional $\tilde{e}$ in \eqref{eq:funct} combines computational ease with a relative flexibility in predicting density dependence of the energy per particle at zero temperature and the momentum dependence of the nucleonic mean fields. It reads
\begin{equation}
\tilde{e} = \sum_X g_X \int \frac{\text{d}{\pmb p}}{(2\pi)^3} \, f_X({\pmb p}) \, \bigg( m_X + \int_0^p \text{d}p' \, v_X^*(p',\rho) \bigg) + \int_0^\rho
\text{d} \rho' \, U_\rho(\rho') \, ,
\label{eq:etilde}
\end{equation}
where the summation is over particle species X. The in-medium particle velocity in the local frame (denoted by the star superscript) depends on the local density and kinematic momentum as
\begin{equation}
v_X^* = \frac{p}{\sqrt{p^2 + m_X^2 \bigg/ \bigg( 1 + c \, \frac{m_N}{m_X} \frac{\rho}{\rho_0} \frac{1}{\big(1+\lambda p^2 /m_X^2\big)^2} \bigg)^2}} \, .
\end{equation}
where m$_\text{X}$ is free mass of particle $X$, and $c$ and $\lambda$, representing the strength and characteristic scale  of the momentum dependence, are adjustable parameters.  In \eqref{eq:etilde}, $U_\rho$ is a density-dependent contribution to the mean-field taken in an empirical form.  The full potential $U(\rho,{\pmb p})$ consists of $U_\rho$ and a  $\delta U_p$ which describes the momentum dependence of the mean field ~\cite{danielewicz2002}:
\begin{equation}
U=U_\rho + \delta U_p = (a \, \rho+b \, \rho^\nu)/[1+(0.4 \, \rho/\rho_0)^{\nu-1}] + \delta U_p \, .
\end{equation}
The constants $a$, $b$, $c$ and $\lambda$ are adjusted to empirical properties of symmetric nuclear matter and characteristics of the momentum dependent mean-field potential (cf. Table~\ref{tab:1}). The functional (\ref{eq:etilde}) parallels the Skyrme density functional (\ref{tdhf}), but contains more elaborate momentum averages than the kinetic energy density used in the Skyrme models. A simple effective mass substitution in the equations of motion, following from the Skyrme functional, would be inadequate in the equations over the range of beam energies explored in heavy-ion collision measurements.

The gradient term $E_{\rm 1}$ allows for the finite range of nuclear forces and its impact, to the lowest order, on the energy and consequently on nuclear densities in the initial state of the reactions. The coefficient  $a_{\rm 1}$  in the gradient term is adjusted to yield realistic diffusivity in the nuclear ground states.  The isospin term  $E_{\rm T}$, dependent on the isospin density  $\rho_\text{T}$, affects both the relative proton and neutron densities in the initial state and the densities reached in the compression stage of a heavy-ion reaction \cite{li2002}. High value of the symmetry energy at supranormal densities may act to reduce the proton-neutron asymmetry at the high densities and, thus, influence the dynamics~\cite{li2002}.

\begin{figure}
 \centering
\includegraphics[width=0.8\textwidth]{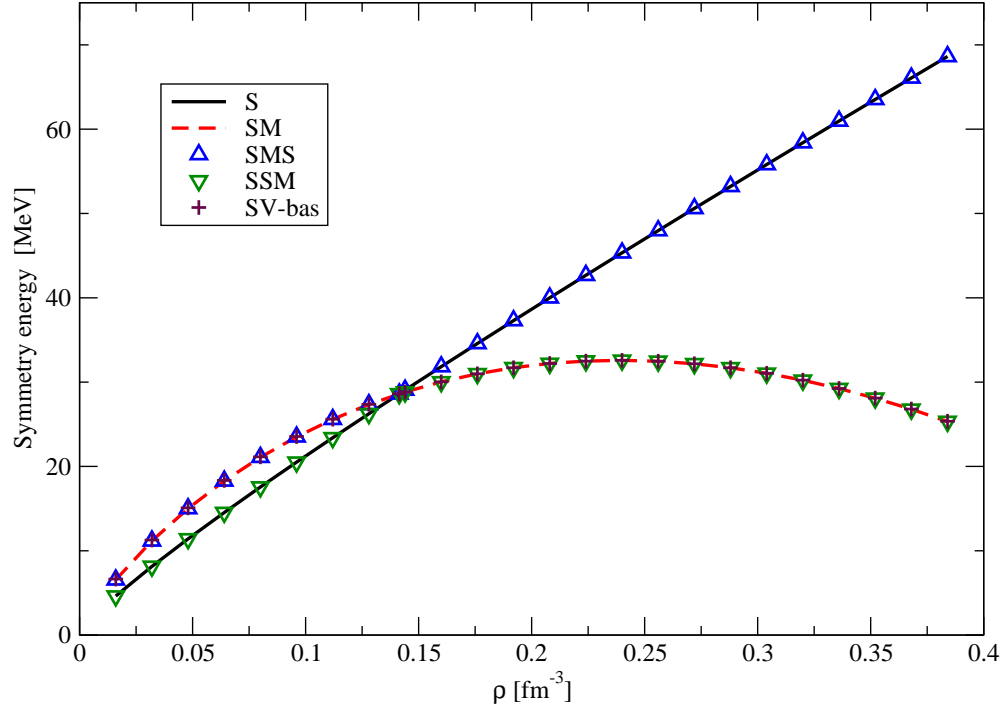}
\caption{\label{fig:1}(Color on-line) Density dependence of the symmetry energy in the models employed in this work. The values of the symmetry energy $S$ and of its slope $L$, at $\rho_0$, are $S(L) =  31.8 (82.8),$ $30. 0 (32.4)$ and $30.2 (32.3) \, \text{MeV}$, for S and SM in pBUU, and for SV-bas, respectively. Models SMS and SSM are added for completeness. }
:\end{figure}

As mentioned in Sec.~\ref{sec:intr}, the density dependence of the S$_\text{int}$ (cf. Eq.~\ref{eq:3}) is an input to the simulation. We employ four different forms of this density dependence. The first model is the conservative choice of a  linear density dependence (labeled S): $S_\text{int} (\rho) = s_{ 0} \, \rho/\rho_0$.  The~second model (labeled SM) mimics  the density dependence predicted by the SV-bas Skyrme interaction \cite{klupfel2009}, used in the TDHF calculation (cf. Sec.~\ref{sec:tdhf}):
$S_\text{int}(\rho) =  s_{\rm 1} \, \rho/\rho_{ 0} - s_{2} \, (\rho/\rho_{0})^{\rm  s_{3}}$ (cf. Table~\ref{tab:1}).
Although these models produce fairly similar values of the symmetry energy at normal density $\sim$$31 \, \text{MeV}$, the slope parameters, $L = 3 \rho_0 \, (\partial S/\partial \rho) |_{\rho_0}$, are significantly different: $L = 85 \, \text{MeV}$ for S  and $L= 31 \, \text{MeV}$ for the SM model.   The hybrid third (SMS) and fourth (SSM) models are constructed using S and SM parameterizations, in order to test  competing impacts of the low-density and high-density behavior of $S_\text{int}(\rho)$.  The third SMS parametrization behaves like SM at low densities and like S at high.  In the SSM parametrization, the order is reversed, i.e.\ S is followed at low and SSM at high densities.  In~each case, the a smooth transition between the S and SM behaviors takes place at a mildly subnormal density $\rho \lesssim \rho_0$, where the S and SM parametrizations cross, cf.~Fig.~\ref{fig:1} where we illustrate the density dependence of the total symmetry energy as predicted by the four models.

The initial state of nuclei entering a reaction is determined by solving the Thomas-Fermi~(TF) equations \cite{danielewicz2000} yielded by the requirement that the total energy (\ref{eq:funct}) is minimal in the ground state . The set of Boltzmann equations (\ref{eq:boltz}) for nucleons, baryon resonances and pions is solved using the test-particle Monte-Carlo method \cite{danielewicz1991}. Given that this work focuses on finding maximum particle densities over position and time, any statistical fluctuations in the Monte-Carlo calculations outside of the model framework may be of concern.  To~minimize possible effects of these fluctuations on deduced maxima, we use a rather large number of test-particles per physical particle \cite{danielewicz2000}, ${\mathcal N}_Q = 3000$.

The pBUU model depends on a number of adjustable parameters as introduced above and  summarized in Table~\ref{tab:1}. The particular combination of parameter values yields fairly conservative properties of nuclear matter, but that parameter set can be varied to explore sensitivity of the progress of a reaction and of reaction observables to the characteristics of the matter.

\subsection{The TDHF model}
\label{sec:tdhf}
The method for treatment of low-energy heavy-ion collisions that currently best combines realism with flexibility and convenience is the Time-Dependent-Hartree-Fock (TDHF) theory, a nonperturbative approach that allows description of multi-nucleon transfer.  At low beam energies, the reaction mechanism of heavy-ion collisions is rather different from that discussed in Sec.~\ref{sec:pBUU}. At energy sufficient to overcome the Coulomb barrier, the colliding system goes through two main phases, fusion and full overlap of the target and projectile, followed by disintegration into two or more fragments.  The typical nucleon mean-free-path at low energies exceeds the size of the composite system. This is in contrast with strongly excited systems produced at higher beam energies, where nucleon mean-free-path shrinks below the system size and continues to decrease as the beam energy increases.

Unlike in high energy collisions when any equilibration has a hard time completing before the reaction ends, in low energy collisions several types of equilibrations may occur, each on a different time scale.  At early stages, a fast charge (chemical) equilibration takes place, which arises from a motion of nucleons near the Fermi surface. The basic mechanism of this equilibration is understood in terms of extension of the single particle motion from one nucleus to another, following the lowering of the potential barrier between the two colliding nuclei after contact.  This process can be very fast, having a typical time scale of $\sim$$10^{\rm -22} \, \text{s} \equiv 30 \, \text{fm}/c$~\cite{iwata2010}. The charge equilibrium is quite important because it prevents production of exotic fragments with extreme proton-neutron asymmetry.  Relatively longer time $\lesssim$$10^{\rm  -21} \, \text{s} \equiv 300 \, \text{fm}/c$ is needed for formation of a composite system in which the single-particle motion becomes synchronized under the influence of the nuclear force to yield collective oscillations. Density equilibration, in which the matter forms a bound state with interior density around the normal value $\rho_{\rm 0}$, may also occur, but is not  well established \IR{and} its time scale is difficult to estimate. It is usually studied together with charge and momentum equilibration.  Momentum equilibrium that represents  a balance between nuclear and Coulomb forces, is expected to take about $10^{\rm -20} \, \text{s}$.  Slower equilibration processes, such as thermal equilibration, and some specific fission and/or decay processes of the composite system, can also take place over times of the order of $10^{\rm -15} \, \text{s}$ in low-energy collisions, but these processes are beyond the scope of this work.

The TDHF dynamics presented in this work has been obtained from Sky3D code designed to solve both static and time-dependent Hartree-Fock equations within a general three-dimensional geometry.  Detailed description of the physics and handling of the code can be found in Ref.~\cite{maruhn2014}.

The model is based on  a Skyrme energy functional of the general form \cite{maruhn2014}
\begin{equation}
E_{\rm tot} =E_\text{kin}+E_\text{Skyrme}+E_{\rm Coul}+E_{\rm pair}+E_{\rm corr} \, ,
\label{tdhf}
\end{equation}
where $E_\text{kin}$ is the total kinetic energy and $E_\text{Skyrme}$  stands for terms related to the Skyrme force. The other terms are additions, not treated self-consistently. They represent the Coulomb and pairing forces and effects of correlations in the mean-field, such as a correction for center of mass motion, a rotational correction for deformed nuclei and a correction for correlations arising from all low-energy quadrupole degrees of freedom in soft nuclei.  In the present work we employ the SV-bas interaction~\cite{klupfel2009} and neglect the pairing and correlation terms in (\ref{tdhf}).  SV-bas is a set of 15 parameters, fitted to 70 independent pieces of data on ground-state binding energies of magic and semi-magic even-even nuclei and to over 75 pieces of experimental data on rms charge and diffraction radii, a neutron skin value and single-particle energies.  In addition, constraints on properties of nuclear matter, namely the value of incompressibility set at $K = 234\, \text{MeV}$, effective mass $m^*/m = 0.9$, the symmetry energy coefficient $a_\text{sym} = 30 \, \text{MeV}$, and the enhancement factor in the Thomas-Reiche-Kuhn sum rule at $\kappa = 0.4$ were imposed in the fit.  For some specific examples of physics application for the SV-bas interaction cf.~Ref.~\cite{iwata2015}.

Both TDHF and  pBUU models are mean-field approaches to particle dynamics in nuclear collisions. They differ in that TDHF is a quantum-mechanical framework utilizing single-particle wavefunctions rather than phase-space probability densities in the semi-classical pBUU.  In TDHF, the initial state of a collision is obtained from static Hartree-Fock (HF) equations as compared to the TF approach in pBUU. A set of time-dependent Schr\"{o}dinger equations is solved self-consistently to obtain wavefunctions describing the time evolution of a colliding system, and yielding spatial proton and neutron densities at any time and location.

The TDHF method is appropriate for simulating low-energy heavy-ion reactions at incident energy up to about $40\, \text{MeV/nucleon}$.  The main reasons for such a limitation is that TDHF does not include any correlations beyond the mean field or N-N collisions which would affect the dynamics. The advantage of the microscopic TDHF approach is that it includes shell effects, most important in low energy reactions, which are not included in the semiclassical pBUU and typically in the molecular dynamics models where, however, N-N collisions are easily incorporated.

\section{Results and Discussion}
\label{sec:res}
In this section we illustrate the main results of the calculations in the whole beam energy range.  We point out similarities and differences in the pBUU and TDHF dynamics across the energy region where both approaches are expected to work.  The proton-neutron asymmetries reached at high densities are examined in order to distinguish contributions of asymmetries inherited from the initial state and coming from high-density behavior of the symmetry energy.

\subsection{Static Densities}
\label{sec:static}

\begin{figure}
 \centering
\includegraphics[width=0.8\textwidth]{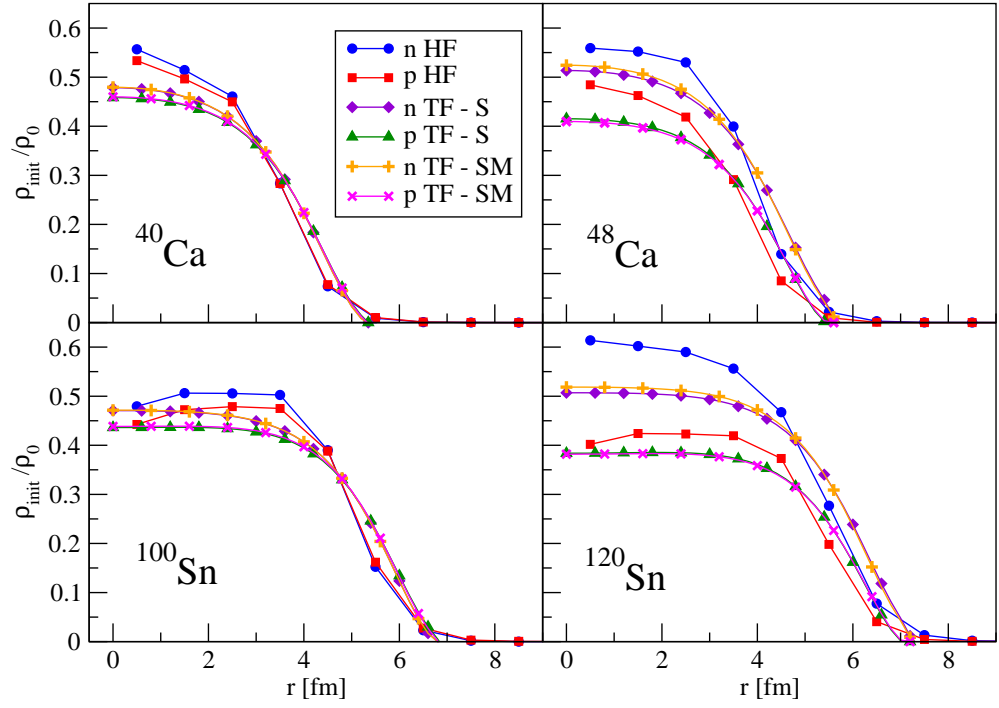}
\caption{\label{fig:2}(Color on-line) Neutron and proton densities as a function of distance from the center of nucleus, for the nuclei considered in the paper, from static HF and TF calculations.}
\end{figure}

Figure \ref{fig:2} shows the density distributions of protons and neutrons, as a function of distance from the center of a nucleus, obtained by solving static TF (in pBUU)  and HF (in TDHF) equations.
In the absence of shell effects within the TF theory, the resulting densities have a smoother dependence on distance and charge and mass number as compared to the HF theory.  Since tunneling effects are not incorporated to the TF theory, the TF densities lack the tails falling off exponentially with distance that are evident in the HF densities.

Following general expectations, the central neutron densities are similar to proton densities in case of the N = Z nuclei, differing by less than 10\%  for both  $^{40}$Ca and $^{100}$Sn in the HF and TF models, cf.~Fig.~2.  The HF model predicts the magnitude of both densities being somewhat higher as compared with the TF model in  $^{40}$Ca.

 In the N $\neq$ Z systems, the HF predictions of magnitude of the central densities are
 also higher than in the~TF model. The central proton-neutron asymmetry is also model dependent. It is about a factor of two bigger for $^{120}$Sn and somewhat smaller for  $^{48}$Ca in HF  as compared to the TF calculation,

These results can be related to differences in physics included in the HF and TF models.  The presence of shell effects in the HF model and its microscopic nature are expected to make the approach more suitable for calculation of static nuclei. Thus, for example, the lack of shell effects and generally the approximate treatment of densities in the TF model can be the reason that, while in $^{120}$Sn the difference between the central proton and neutron densities is higher in HF than~TF, in $^{48}$Ca the differences are reversed, see Fig.~2. The shell effects influence also another significant factor that plays a role, the density distributions at the surface as compared to the center of a nucleus.  In the lighter Ca systems the ratio of  surface to volume is larger that in the Sn systems. As the symmetry energy is lower at lower densities,  the surface areas are energetically favorable for development of relative asymmetry as compared to the center where the density is higher. Thus the proton-neutron imbalance at the center is expected to be lower in Ca than in Sn nuclei. As discussed more in detail  in Sec.~\ref{sec:pnasym}, the static proton-neutron asymmetry and its development at subnormal densities affect significantly the asymmetry at supranormal densities and, thus, has to be determined as well as possible in order to make trustworthy reaction predictions.

The choice of a symmetry energy model from our pool affects very little the central proton and neutron densities in the TF calculations. We illustrate this effect showing results for~S and SM in Fig.~\ref{fig:2}. The results for SSM and SMS models are not included as they are practically identical to those obtained with S and SM models, respectively. This is to be expected as there is very little difference within the pairs of the models at sub-saturation densities (cf.~Fig.~\ref{fig:1})

\subsection{Time evolution of a collision: contrasting approaches, system sizes and energies}
\label{sec:time}

\begin{figure}
 \centering
\includegraphics[width=\textwidth]{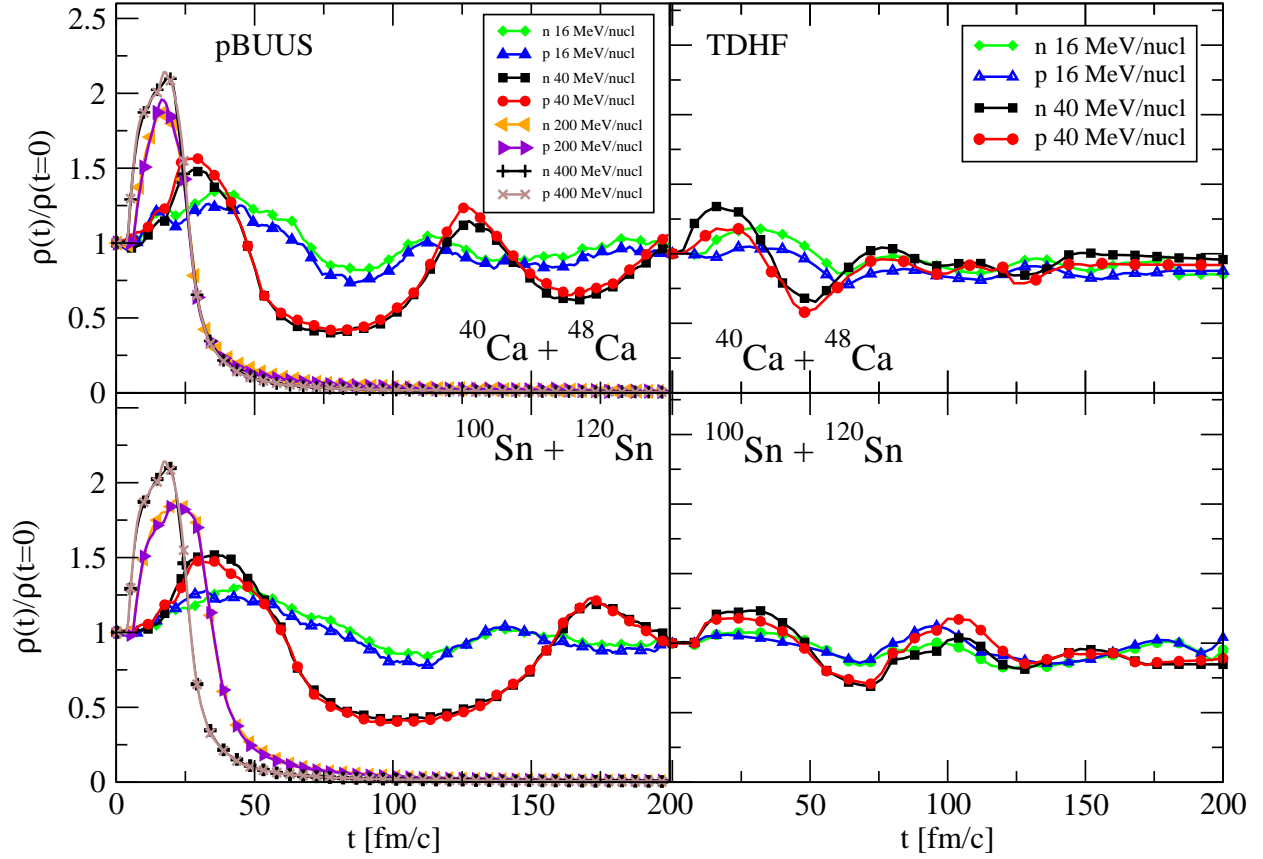}
\caption{\label{fig:3}(Color on-line) Time evolution of maximal proton and neutron densities normalized to their static value,  in asymmetric collisions, $^{\rm 40}$Ca+$^{\rm 48}$Ca and $^{\rm 100}$Sn+$^{\rm 120}$Sn, at different incident energies, within pBUU dynamics for symmetry energy S (left panels) and in TDHF dynamics (right panels). }
\end{figure}

Detailed study of time evolution of proton and neutron densities as a function of beam energy is essential to our particular goal of extracting maximal values of the densities during a central collision.  Two representative cases of time evolution are illustrated in Fig.~\ref{fig:3} for asymmetric \ $^{\rm 40}$Ca + $^{\rm 48}$Ca and $^{\rm 100}$Sn +  $^{\rm 120}$Sn projectile+target systems. Results yielded by the pBUU dynamics with the S model for the symmetry energy, for beam energy  $16 \leq E_{\rm beam} \leq 400 \, \text{MeV/nucl}$ are complemented by the outcome of the TDHF calculation for  $16 \leq E_{\rm beam} \leq 40 \, \text{MeV/nucl}$, the beam energy region where the TDHF model is applicable.

The time-dependence of maximal densities in Fig.~\ref{fig:3} exhibits some common features characteristic for all the systems we investigate, irrespectively of the approach.  First, the neutron and proton density tends to maximize at about the same time, i.e. the maximum compression of neutrons and proton matter is synchronized.  Second, the time elapsed since early contact between the colliding nuclei till the absolute maximum is reached is beam energy dependent, dropping as the energy is increased.

In more energetic collisions, modeled only in pBUU, the maximal density drops in a rather dramatic fashion after the maximum compression has been reached.  This drop corresponds to a system vaporization. The times for maximal compression and vaporization change little for the Ca systems between $E_\text{beam} = 200$ and $400 \, \text{MeV/nucl}$.  In the Sn systems more noticeable delays develop with energy.

\begin{figure}
 \centering
\includegraphics[width=\textwidth]{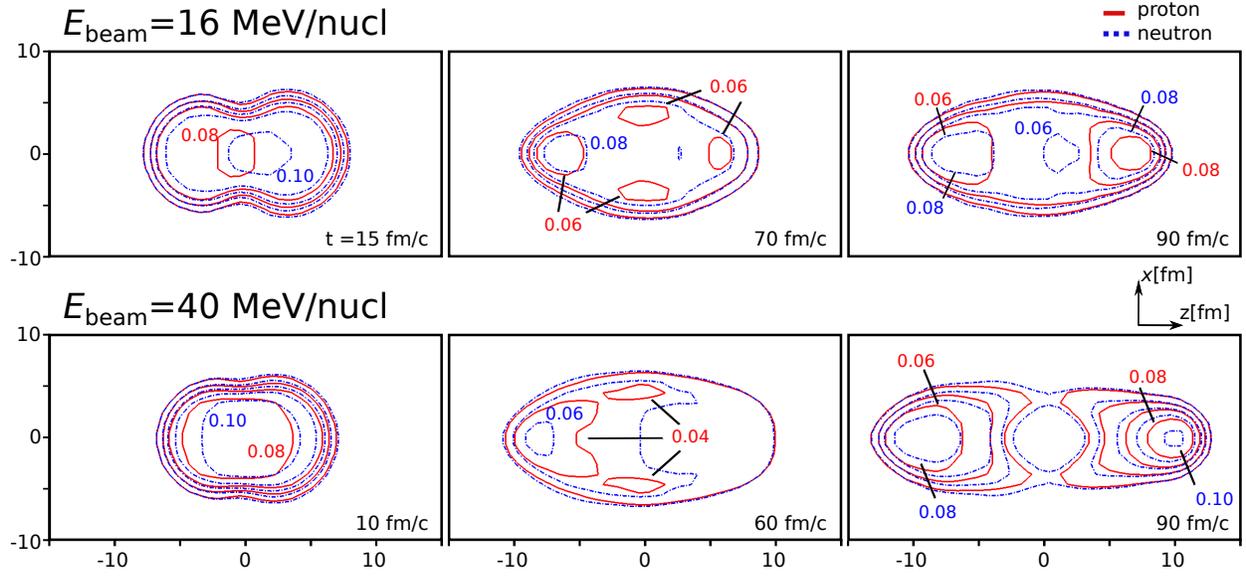}
\caption{\label{fig:4}(Color on-line) Contour plots of neutron (dashed lines) and proton density (solid lines) in head-on $^{120}$Sn + $^{100}$Sn collisions at $16\,\text{MeV/nucl}$ (top) and $40\,\text{MeV/nucl}$, at different times. The~horizontal axis is the collision axis.}
\end{figure}

Results at $E_{\rm beam} \leq  40 \, \text{MeV/nucl}$ offer not only exploration of the lower beam energy region but also a comparison of the pBUU and TDHF model predictions. As illustrated in Fig.~\ref{fig:3}, in both the Ca and Sn systems the first maximum, typically absolute, is followed by a series of oscillations with amplitudes decreasing faster with time in pBUU than in TDHF.
The nature of these oscillations in TDHF is illustrated in Fig.~\ref{fig:4}, which displays contour plots of proton and neutron densities at sample times in the  $^{100}$Sn+$^{120}$Sn collisions at the energies of 16 and $40 \, \text{MeV/nucl}$. At lowest energies, the oscillations represent motion of a fused system.  At somewhat higher energies, the systems tend to fission after reaching the maximum compression and the oscillations represent motion of separated fission fragments.

The out-of-phase oscillations in neutron and proton densities, particularly visible in TDHF, are related to waves propagating across the colliding system that are fast compared to the overall system evolution.  In a projectile-target asymmetric system, the oscillation starts with a charge equilibrating wave propagating across the fusing nuclei. This wave is combined with a spontaneous dipole oscillation along the direction of the reaction that adds, with an \textit{ad hoc} phase, to the existing proton and neutron density differences.

The shape relaxation, combined with Coulomb repulsion, enhances multipolarity of the isovector waves, as demonstrated in Fig.~\ref{fig:4}. Here we see fission, following the early fusion, traps instantaneous asymmetries at both ends of the system within the separating fragments which subsequently re-contract.  This relatively complicated and weakly damped isovector dynamics in TDHF leads to effects such as prediction of a stronger compression of protons than neutrons in the second peak of the oscillation (cf. Fig.~\ref{fig:3}).

Comparing further the Ca and Sn systems in the pBUU and TDHF models, relaxation after the first compression takes longer in pBUU than in TDHF. These times are almost the same for Ca and Sn systems in TDHF, but in pBUU the Ca system relaxes faster than the Sn system. Although the overall pattern of the density evolution is not very different between the approaches at $E_{\rm beam} = 16 \, \text{MeV/nucl}$, for both Ca and Sn, there is a markable difference in predictions of the TDHF and pBUU models for $E_{\rm beam} = 40 \, \text{MeV/nucl}$, both in amplitudes of the first compression and the onset of oscillations.  This effect may be related to the collisional equilibration taking place in pBUU but not TDHF and to overall stronger damping of collective waves in pBUU than TDHF.

\subsection{Maximum densities and the proton-neutron asymmetry}
\label{sec:pnasym}
Using the maximal densities $\rho_p^\text{max}$ and $\rho_n^\text{max}$, extracted from data on  time evolution of a collision as detailed in Sec.~\ref{sec:time}, we construct the asymmetry $\delta = (\rho_n^\text{max} - \rho_p^\text{max})/(\rho_n^\text{max} + \rho_p^\text{max})$ as a function of system and beam energy. Although, strictly speaking, the maximal densities may not be reached at exactly the same time, or at the same location,  we find that $\delta$ gives a good overall representation of the asymmetry in the densest matter in a collision, even in systems with different projectile and target initial asymmetries.

\begin{figure}
 \centering
\includegraphics[width=0.8\textwidth]{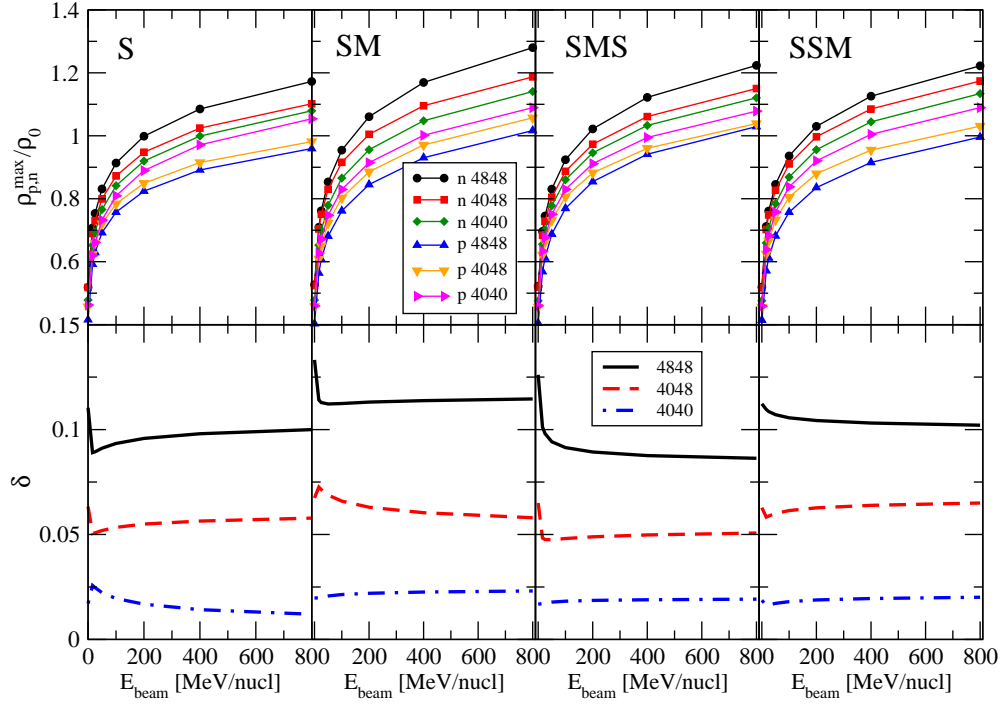}
\caption{\label{fig:5}(Color on-line) Maximum neutron and proton densities in units of $\rho_{\rm 0}$ (top panels)  and the corresponding values of the asymmetry $\delta$ (bottom panels), vs beam energy for Ca systems, yielded in the pBUU simulations with different indicated forms of symmetry energy.}
\end{figure}

\begin{figure}
 \centering
\includegraphics[width=0.8\textwidth]{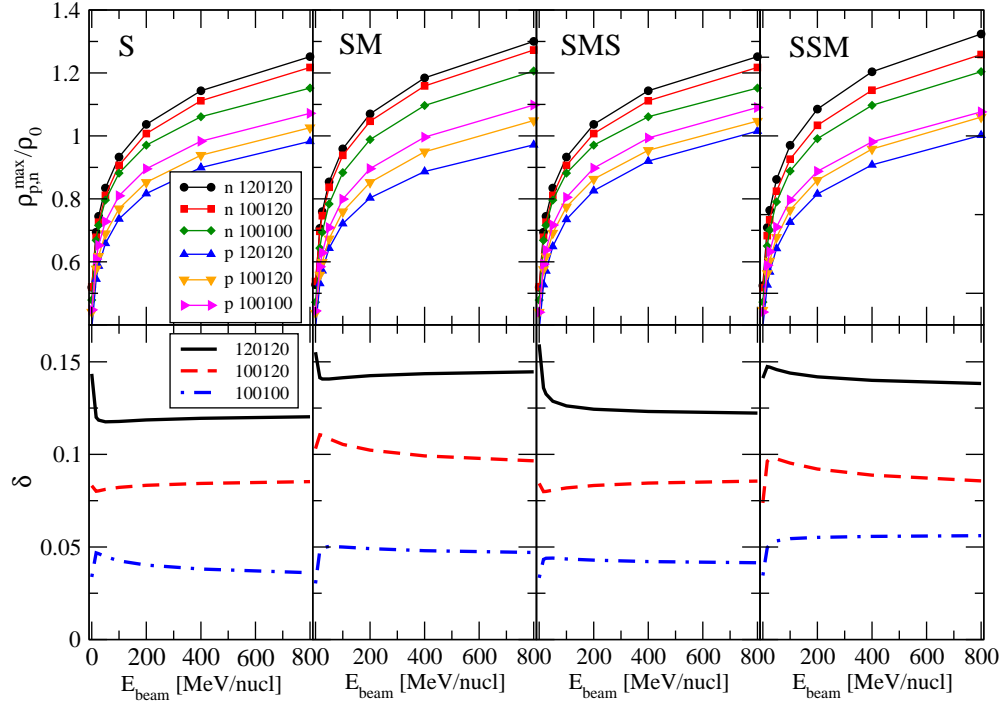}
\caption{\label{fig:6}(Color on-line) The same as Fig.~\ref{fig:4}, but for Sn systems.}
\end{figure}

\begin{figure}
 \centering
\includegraphics[width=0.8\textwidth]{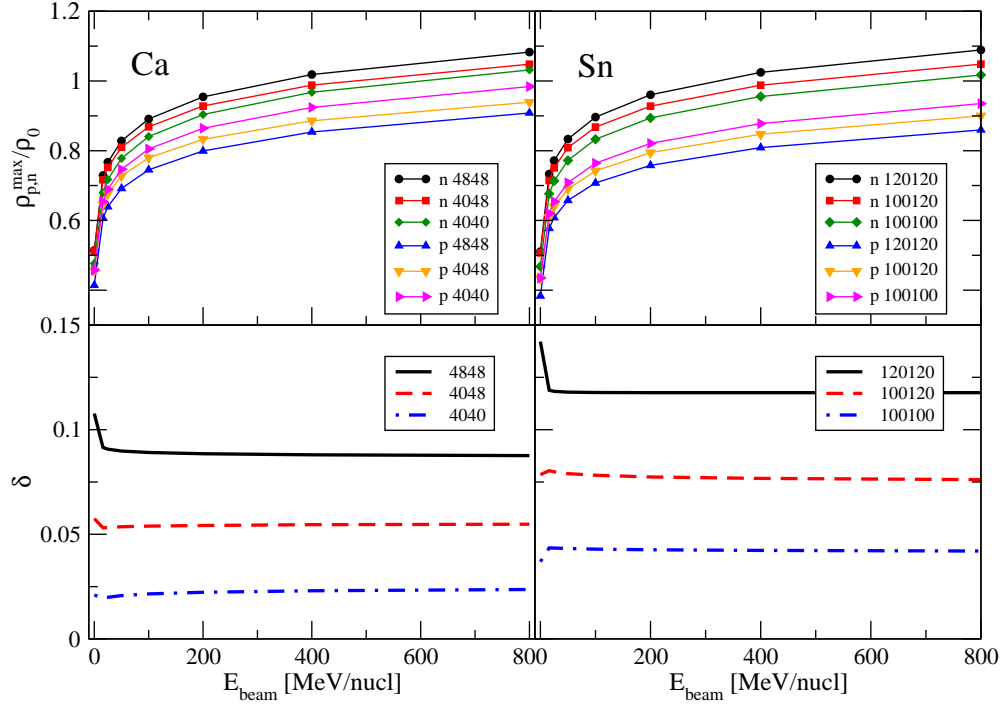}
\caption{\label{fig:7} (Color on-line) Maximum neutron and proton densities in units of $\rho_{\rm 0}$ (top panels) and corresponding values of asymmetry $\delta$ (bottom panels), vs beam energy for Ca (left) and Sn (right) systems as calculated in Vlasov simulations with symmetry energy model S.}
\end{figure}

\begin{figure}
 \centering
\includegraphics[width=0.8\textwidth]{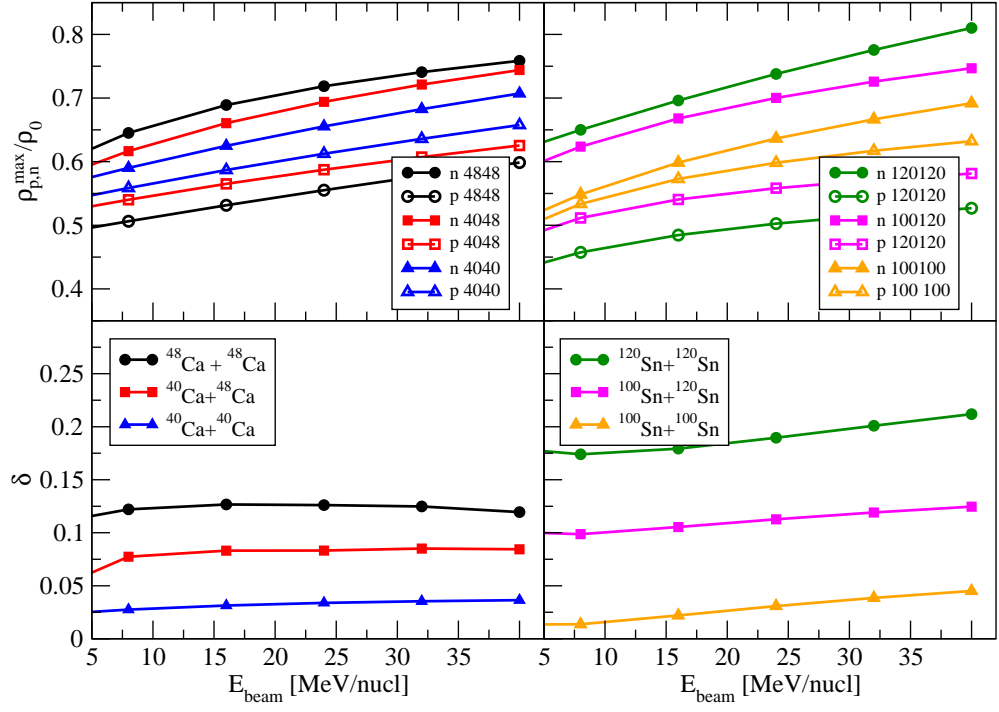}
\caption{\label{fig:8} (Color on-line) Maximum neutron and proton densities in units of $\rho_{\rm 0}$ (top panels) and the corresponding values of asymmetry $\delta$ (bottom panels), vs beam energy for Ca (left) and Sn (right), from TDHF simulations.}
\end{figure}

The proton and neutron densities fluctuate as a function of  beam energy in pBUU as  a consequence of the Monte-Carlo nature of the calculations and due to the propagating waves with a strong isovector component in TDHF. The fluctuations can obscure gradual changes, particularly in a differential quantity such as the asymmetry $\delta$. To circumvent the issue of fluctuations, in addition to extracting individual values of the maximum densities, we approximate them at each beam energy $x$ using a smoothing function  $y = a_{\rm 0}+a_{\rm_1}\log(1+a_{\rm 2} x)$. These interpolated densities, shown in Figs.~\ref{fig:5}--\ref{fig:7} for pBUU and in Fig.~\ref{fig:8} for TDHF,  are then used to calculate $\delta$. The parameters $a_{\rm 0}$, $a_{\rm_1}$ and $a_{\rm 2}$, summarized in the Supplemental Material, show reasonable consistency in all related cases. It follows that the interpolation function is an adequate representation of the raw data. We note that the error in the fitting is within a few percent in all cases and does not affect conclusions of the analysis.

Figures~\ref{fig:5}--\ref{fig:7} illustrate the evolution of the interpolated maximum proton and neutron densities with increasing beam energy for the Ca and Sn systems, respectively, as simulated in pBUU, using four models of the density dependence of the symmetry energy S, SM, SMS and SSM. The results indicate that $\rho_{\rm n}^{\rm max}$ does not exceed $\sim$1.3$\times\rho_{\rm 0}$ for Sn systems, with the $\rho_{\rm p}^{\rm max}$ being below $\sim$1.1$\times\rho_{\rm 0}$. The results for the lighter Ca systems are marginally lower. In the Vlasov approximation (cf.~Fig.~\ref{fig:7}), in which the N-N collisions are eliminated, the maximum densities are lower. The highest maximum particle number density, as resulting from all models used in this work for  Ca and Sn systems and beam energies, is $\sim$2.5$\times\rho_{\rm 0}$ at 800 MeV/nucl.

The total nucleon densities normalized to the initial static values (compression ratios) are given in Tables \ref{tab:2} and \ref{tab:3} for pBUU and TDHF, respectively.  The ratios generally rise with increasing beam energy. Interestingly, the compression ratios depend on the size of a~system. Within the elements, the pBUU model predicts the highest compression ratios for the heaviest (symmetric) Ca and Sn systems, respectively, cf.~Table~\ref{tab:2}. The TDHF model yields the same result for Ca, but in the Sn systems the situation is reversed and  $^{\rm 100}$Sn +  $^{\rm 100}$Sn combination yields the highest compression.  This is likely related to a strong shell effect, enhancing initial density in $^{120}$Sn, that apparently dissolves during the collision, but is used in the normalization. In~the region of overlap of the applicability of pBUU and TDHF, we consistently find higher compression ratios in the pBUU model.  The difference is even more obvious from comparison of TDHF to pBUU in the Vlasov mode, conceptually closer to TDHF, as neither of these models includes elementary N-N collisions.  It seems that the N-N collisions are detrimental to compression at lower energies, but they enhance it at higher energies, cf.~Table~\ref{tab:2}.
 The TDHF model yields systematically lower compression in the beam energy region below $40 \, \text{MeV/nucl}$, than the pBUU model in the same region, reaching maximum compression ratio $\sim$1.4 at 40 MeV/nucl for Sn and $\sim$1.3 for the Ca systems (cf.~Table~\ref{tab:3} and Fig~\ref{fig:8}).
Higher compression ratios are predicted above the beam energy  $50 \, \text{MeV/nucl}$, reaching $\sim 2.5$ at $800 \, \text{MeV/nucl}$ in the pBUU model with N-N collisions included.

A remarkable feature of the asymmetries $\delta$, shown in the bottom panels of \mbox{Figs.~\ref{fig:5}--\ref{fig:7}},  is that they evolve with the beam energy more rapidly at low energies than at high energies.  This effect may be understood in terms of the pace of the dynamics.  At lower beam energies, the dynamics is more adiabatic, with density profiles and asymmetry values undergoing broader adjustments before maximal densities being reached.  At high energies, the central region of the system becomes compressed in a shock manner and the asymmetry represents closer the matter that suddenly has fallen into that region.  In the Ca systems, a significant portion of proton-neutron imbalance is pushed out to the surface, so asymmetries at maximum compression are much reduced, almost by a factor of two, compared to the total asymmetry expected from the total number of protons and neutron in the system. The~asymmetries at maximum densities in the Sn systems are higher, both because of the lower surface area to volume ratio and as a consequence of stronger Coulomb repulsion, disfavoring protons in the central region of the system.  Comparing results with and without N-N collisions (in the Vlasov dynamics), in Figs.~\ref{fig:5}--\ref{fig:7}, we see that when the N-N collisions are active, they help to trap the asymmetry in the high-density regions.  The N-N collisions have also a different impact on asymmetry in dependence on the size of a system. The mean-free-path between N-N collisions is shorter, relative to system size, for Sn than for Ca, making the collision more effective in trapping the isospin excess in high density regions, thus increasing the asymmetry coming from the initial proton-neutron distribution and Coulomb repulsion.

In comparing asymmetries $\delta$ reached in TDHF and pBUU, in the overlap region for the approaches, Fig.~\ref{fig:8} and \mbox{Figs.~\ref{fig:5}--\ref{fig:7}}, it is observed that the asymmetry for maximal densities is generally higher in TDHF than in pBUU, particularly at low energy.  This effect seems to be related to the undamped waves in TDHF that, combined with fission, increase asymmetry in one fragment at the cost of the other.  The fragment separation enhances the maximal neutron density less than the maximal proton density, due to the increased Coulomb repulsion in the fragment with reduced asymmetry, i.e.\ extra charge number per neutron. With increasing incident energy the predictions of TDHF and of pBUU for the SM energy parametrization get closer to each other.

The apparent impact of the sudden dive of matter into a high density region, on the high-density proton-neutron asymmetry in collisions at high energy, raises questions regarding the feasibility of probing the high-density symmetry energy in the laboratory.
Comparing the high-energy asymmetries in the panels labeled S and SM in Figs.~\ref{fig:5} and~\ref{fig:6}, it could be argued that the simulation with higher symmetry energy at high density, i.e.~S model cf.~Fig.~\ref{fig:1}, yields lower asymmetries as the excess neutrons are expelled from the high-density region~\cite{li2002}. However, it could  also be argued that the difference in the high-density asymmetries observed at high beam energies, resulting from S and SM calculations, stems exclusively from the difference in interior asymmetries within the nuclei at the start of a~collision, dooming the possibility of testing the high-density symmetry energy in heavy-ion collisions.

 In order to examine a possible impact the high-density symmetry energy may have, we constructed the hybrid symmetry-energy models, SSM and SMS, for which the initial states are practically identical with the S and SM models, respectively, but the high-density behavior of the symmetry energy is swapped in-between the models.  Comparing S with SSM and SM with SMS for Ca and, even more importantly, for Sn, we see that the high-density behavior of the symmetry energy does make a difference.  If the symmetry energy is switched at high density from stiff (raising fast with increasing density) to soft raising slowly with increasing density), i.e.\ S to SSM, the high-energy asymmetry increases. On the other hand, if the models are reversed,  i.e.\ SM to SMS, the high-energy asymmetry decreases.

It seems that the SMS model is more effective in loosing asymmetry than SSM model, particularly in the lighter Ca systems where the reaction lasts for a shorter time.  In other words, it is easier to lose asymmetry than to reconstitute it after it has been depleted in the dynamics when the high-density symmetry energy is stiff and \textit{vice versa}.  Overall, the experimentation with the symmetry energy models shows that both, the initial distribution of asymmetry in the nuclei and the high-density behavior of the symmetry energy impact the high-density asymmetry.  To reveal the effect of the initial distribution of asymmetry on the high-density behavior of symmetry energy, the static nucleon distributions in the target and projectile nuclei must be modeled consistently in the simulations.

For the most proton-neutron asymmetric systems,  $^{48}$Ca + $^{48}$Ca and $^{120}$Sn + $^{120}$Sn, we find changes in the high-density asymmetry of up to $\sim 20$\% in the lighter system and up to $\sim 15$\% in the heavier when the symmetry energy is varied, cf.~Figs.~\ref{fig:5} and \ref{fig:6}.  Paradoxically, the higher sensitivity to symmetry energy in the lighter system seems to be related  to the combined ratios of surface thickness in the initial state and the mean-free-path to system size in the dynamics. Importantly if the nuclear systems were very large, the asymmetry at high density would neither change due to low- nor high-density behavior of the symmetry energy.  This is because the neutron-proton imbalance would have practically nowhere to move from the interior of matter neither in the initial nor in the compressed state.  With this, while the general expectation is that heavier systems are better suited for testing EOS, in the particular case of supranormal symmetry energy, lighter systems may be more suitable, when the dedicated observables are tied to the value of asymmetry reached at high densities.

\section{Conclusions and outlook}
\label{sec:concl}

We presented results of the first systematic examination of maximal values and other details in neutron and proton distributions, within the initial state and throughout the course of head-on heavy-ion collisions across the beam energy range of up to $800 \, \text{MeV/nucl}$, in two common simulation approaches, TDHF and pBUU.  The projectile-target symmetric systems $^\text{40}$Ca +$^\text{40}$Ca,  $^\text{48}$Ca +$^\text{48}$Ca, $^\text{100}$Sn +$^\text{100}$Sn and $^\text{120}$Sn + $^\text{120}$Sn were selected for this study, as well as the projectile-target asymmetric systems $^\text{40}$Ca +$^\text{48}$Ca and $^\text{100}$Sn +$^\text{120}$Sn.  Within the pBUU simulations, two empirical forms of symmetry energy, characterized by $L=85 \, \text{MeV}$, termed S, and $L=31 \, \text{MeV}$, termed SM, and two hybrid forms, termed SSM and SMS were employed. Simulations with suppressed collisions, i.e.\ in the Vlasov mode were also performed for completeness. TDHF calculations were carried out in the beam energy range of up to $40 \, \text{MeV/nucl}$, with the SV-bas Skyrme parametrization that yields the symmetry energy close to that in the SM parametrization within pBUU.  Maximum proton and neutron densities $\rho_{\rm p}^{\rm max}$ and $\rho_{\rm n}^{\rm max}$, reached in the course of the collisions, were identified from analysis of the time evolution of  $\rho_{\rm p}$ and  $\rho_{\rm n}$ during the course of the collision and used to yield the proton-neutron asymmetry $\delta = (\rho_{\rm n}^{\rm max} - \rho_{\rm p}^{\rm max})/ (\rho_{\rm n}^{\rm max} + \rho_{\rm p}^{\rm max})$ as a~function of beam energy and all target/projectile combinations considered in this work.

In TDHF, the initial densities, as calculated the HF model, are characterized by significant shell oscillations unlike the TF starting densities for pBUU.  The oscillations, combined at low energies with weakly damped collective waves with significant isovector components, enhance density variations across the system and time. When a colliding system fissions, the variations may become trapped in opposite fragments, leading to a larger proton-neutron imbalance than in the system before fission. In combination with Coulomb effects, this effect enhances the proton-neutron asymmetry at the maximal recorded densities. As the beam energy increases, the impact of the initial-state shell effects decreases and the TDHF results become similar to pBUU.

The pBUU simulations indicate that the maximum total densities reached in reactions studied in this work are $\sim 2.5 \, \rho_0$ ($0.4 \, \text{fm}^{\text -3}$) are not significantly dependent on the initial state, beam energy, system size and a model of the symmetry energy.
However, the proton-neutron asymmetry at maximum densities does depend on these parameters. The simulation in this work indicates without any doubt that the initial asymmetry of all systems studied plays an important role in a reaction. The asymmetry decreases after the impact, changes slowly in the course of a collision as a function of beam energy, but never exceeds the original value.  For example,  for the most neutron-proton asymmetric systems, i.e.\ $^{48}$Ca + $^{48}$Ca and $^{120}$Sn + $^{120}$Sn, both with the overall initial asymmetry of 0.17, we find high-density asymmetries at high beam energies in the range of (0.09--0.12) for the lighter system and (0.12--0.15) for the heavier, depending on the symmetry energy model employed.  It has become clear that in predicting the impact of the density dependence of symmetry energy on high-density asymmetry, the initial state must be modeled consistently with the dynamics that leads to development of the compressed matter. If the rate of change of the initial asymmetry could be measured as a function of the beam energy, it would give information of the high-density behavior of the symmetry energy.  When changing the symmetry energy, we find somewhat more impact on the high density asymmetry in the lighter asymmetric systems than heavier.

To summarize, the main outcome of our work is the identification of importance of the sub-normal density dependence of the symmetry energy in determination of its form in supra-normal density. There is an effort in the community to learn about the supra-normal symmetry energy, with different variants explored in theoretical simulations.  One practical issue is that changes in the symmetry energy model in the supra-normal region, smoothly matched to a model in the sub-normal region, inevitably affects the initial state of the collision. Our simulation reveals that only about half of the proton-neutron asymmetry at super-normal densities comes directly from behavior of the supra-normal symmetry energy and the remaining half has its origin in changes in the initial state. This result puts to forefront the need of a proper initialization of the nuclei, but also brings up the question of the shell effects that impact initial proton and neutron densities, but cannot be consistently incorporated into semiclassical transport.  In exploring the density dependence of the symmetry energy in central reactions, lighter systems may be as useful as heavier.

\begin{acknowledgments}
J.R.S.\ acknowledges, with pleasure, the support from the Japanese Grant-In-Aid for Scientific Research: Nuclear Matter in Neutron Stars Investigated by Experiments and Astrophysical Observations: Sub-Group B01 (Tetsuya Murakami), to attend the Workshop Nuclear Symmetry-Energy and Nucleus-Nucleus Collision Simulation (RIBF-ULIC-miniWS027), where the idea of this paper was initiated. Y.I.\ was supported by MEXT SPIRE and JICFuS.  The TDHF numerical computations were carried out at the Computer Facility of the Yukawa Institute at Kyoto University.  P.D.\ acknowledges support from the U.S.\ National Science Foundation under Grant PHY-1403906 and from JUSEIPEN (Japan-U.S.\ Experimental Institute for Physics with Exotic Beams) under a Grant from the U.S.\ Department of Energy.
\end{acknowledgments}

\clearpage
\vskip 2cm

\clearpage
\vskip 2cm

\clearpage
\vskip 2cm

\clearpage
\vskip 2cm

\clearpage
\vskip 2cm

\clearpage
\vskip 2cm

\clearpage
\vskip 2cm

\clearpage
\vskip 2cm

\clearpage

\begin{table}
\caption{Adjustable parameters in the pBUU and TDHF models.}
\label{tab:1}
\begin{tabular}{||c|l|c|c||}
\hline
\hline
Model & Parameter Group & Parameter & Value \\
\hline
\hline
pBUU  & Gradient Term in Energy $E_1$ & $a_1$ & $21.4 \, \text{MeV}\, \text{fm}^2$  \\
\cline{2-4}
      & Density-Dependent & $a$ & $209.791 \, \text{MeV}$ \\
      &   contribution to Mean-Field $U_\rho$ & $b$ & $69.7571 \, \text{MeV}$ \\
      &                               & $\nu$ & 1.46226 \\
      &                               & $\rho_0$ & $0.160 \, \text{fm}^{-3}$ \\
      &                               &  $K$  & $210 \, \text{MeV}$  \\
\cline{2-4}
      & Momentum-Dependent &  $c$ & $0.645700$ \\
      &  contribution to Mean Field $\delta U_p$ & $\lambda$ & 0.954605 \\
      &                                          & $m^*/m$   &  0.70  \\
\cline{2-4}
      & Interaction Contribution  &  $s_0$   &  $20.0 \, \text{MeV}$ \\
      & to Symmetry Energy $S_\text{int}$ &  $s_1$   &  $50.1 \, \text{MeV}$ \\
      &                                   &  $s_2$   &  $31.9 \, \text{MeV}$  \\
      &                                   &  $s_3$   &  $1.47$ \\
\cline{2-4}
      & In-Medium Cross Section & $\nu_\text{cs}$  &    0.667 \\
      & Eqs.~(11)-(12) of Ref.~\cite{danielewicz2009a} &    &   \\
\cline{2-4}
      &  Monte-Carlo Integration                 & ${\mathcal N}_Q$  & 3000\\
\hline
\hline
TDHF Model & SV-bas & \multicolumn{2}{l||}{11 parameters in Ref.~\cite{klupfel2009}} \\
\hline
\hline
\end{tabular}
\end{table}

\clearpage

\begin{landscape}
\begin{center}
\setlength{\LTcapwidth}{6.1in}
\begin{longtable}
               {||c | c c c c c c c c c c c c c c c||}
\caption[]{Ratio of maximal nucleon density reached at different beam energies $E_\text{beam}$, to maximal density in the initial state, for the Sn and Ca systems, as predicted in the pBUU, with the symmetry energy  models S, SM, SMS and SSM and in the Vlasov mode (V) with the symmetry energy model S.  The columns are labeled with letters representing the version of the model, followed by mass numbers of the target and projectile combinations. \label{tab:2} }
\\
\hline \hline
& & & & & & & & & & & & & & & \phantom{X} \\
E$_{\rm beam}$ &  & S & &  & SM & & & SMS & & & SSM & & & V & \\[.5ex]
MeV/nucl & 120120& 100120& 100100 &   120120& 100120& 100100 & 120120& 100120& 100100 & 120120& 100120& 100100 & 120120& 100120& 100100  \\[.5ex] \hline
& & & & & & & & & & & & & & & \phantom{X} \\
0		&1.00	&1.00	&1.00	&1.00	&1.00	&1.00   &1.00 &1.00 &1.00  &1.00  &1.00  &1.00    &1.00	&1.00	&1.00  \\[.5ex]
16		&1.37	&1.30	&1.38	&1.36	&1.29	&1.34   &1.34 &1.32 &1.36  &1.35  &1.29  &1.36    &1.47	&1.48	&1.44  \\[.5ex]
25		&1.47	&1.39	&1.48	&1.46	&1.38	&1.44   &1.45 &1.42 &1.47  &1.45  &1.39  &1.46    &1.54	&1.56	&1.51 \\[.5ex]
50		&1.64	&1.55	&1.65	&1.64	&1.55	&1.63   &1.64 &1.60 &1.65  &1.64  &1.56  &1.64    &1.67	&1.68	&1.64\\[.5ex]
100		&1.84	&1.72	&1.83	&1.84	&1.74	&1.84   &1.85 &1.79 &1.85  &1.85  &1.75  &1.85    &1.79	&1.81	&1.77\\[.5ex]
200		&2.04	&1.91	&2.01	&2.05	&1.95	&2.06   &2.08 &1.99 &2.06  &2.07  &1.96  &2.06    &1.92	&1.93	&1.90\\[.5ex]
400		&2.25	&2.10	&2.21	&2.27	&2.16	&2.29   &2.31 &2.20 &2.28  &2.30  &2.18  &2.28    &2.05	&2.06	&2.03\\[.5ex]
800		&2.46	&2.30	&2.40	&2.49	&2.38	&2.52   &2.55 &2.42 &2.50  &2.53  &2.40  &2.50    &2.18	&2.19	&2.16\\[.5ex]  \hline
& & & & & & & & & & & & & & & \phantom{X} \\
	        &4848	&4048	&4040	&4848	&4048	&4040  &4848 &4048  &4040 &4848 &4048 &4040 &4848	&4048	&4040 \\[.5ex] \hline
& & & & & & & & & & & & & & & \phantom{X} \\
0	        &1.00	&1.00	&1.00	&1.00	&1.00	&1.00   &1.00 &1.00  &1.00  &1.00  &1.00  &1.00   &1.00	&1.00	&1.00\\[.5ex]
16     	&1.39	&1.35	&1.35	&1.37	&1.34	&1.37   &1.37 &1.34  &1.38  &1.37  &1.36  &1.39   &1.44	&1.41	&1.43\\[.5ex]
25	        &1.48	&1.43	&1.44	&1.47	&1.43	&1.46   &1.46 &1.43  &1.47  &1.47  &1.45  &1.49   &1.52	&1.48	&1.51\\[.5ex]
50	        &1.63	&1.56	&1.59	&1.65	&1.58	&1.63   &1.64 &1.58  &1.63  &1.64  &1.60  &1.65   &1.64	&1.59	&1.63\\[.5ex]
100	        &1.79	&1.70	&1.75	&1.85	&1.75	&1.81   &1.83 &1.74  &1.81  &1.81  &1.76  &1.83   &1.76	&1.70	&1.76\\[.5ex]
200	        &1.95	&1.85	&1.92	&2.05	&1.92	&1.99   &2.02 &1.90  &1.98  &2.00  &1.93  &2.01   &1.89	&1.82	&1.89\\[.5ex]
400	        &2.12	&2.00	&2.09	&2.26	&2.10	&2.19   &2.23 &2.07  &2.17  &2.19  &2.10  &2.19   &2.02	&1.94	&2.03\\[.5ex]
800	        &2.28	&2.14	&2.26	&2.47	&2.28	&2.38   &2.43 &2.24  &2.35  &2.37  &2.27  &2.38   &2.15	&2.05	&2.16\\[.5ex] \hline \hline
\end{longtable}
\setlength{\LTcapwidth}{4in}
\end{center}
\end{landscape}

\clearpage

\begin{table}\caption{\label{tab:3} Ratio of maximal nucleon density reached at different beam energies $E_\text{beam}$, to maximal density in the initial state, for the Sn and Ca systems, as predicted by the TDHF model.  The~columns are labeled with mass numbers of the target and projectile combinations.}
\vspace{10pt}
\begin{tabular}{||c||cccccc||}\hline \hline
 $E_\text{beam}$		&4848	&4048	&4040	&120120 &100120   &100100   \\
 {MeV/nucl}         &          &    &     &     &          &   \\
 \hline  \hline
0		&1.00	&1.00	&1.00	&1.00	&1.00	&1.00\\
4		&1.07	&1.05	&1.03	&1.06	&1.11	&1.11\\
8		&1.11	&1.09	&1.07	&1.11	&1.16	&1.18\\
16		&1.18	&1.15	&1.12	&1.18	&1.24	&1.28\\
24		&1.23	&1.20	&1.18	&1.24	&1.29	&1.35\\
32		&1.27	&1.25	&1.22	&1.29	&1.33	&1.40\\
40		&1.31	&1.29	&1.27	&1.34	&1.36	&1.44\\   \hline \hline
\end{tabular}
\end{table}

\end{document}